\documentstyle[aps,prl,amsfonts,amssymb,epsf,special]{revtex}

\begin{document}

\draft { \title{Dephasing in Metals by Two-Level Systems in the
    2-Channel-Kondo Regime} \author{${}^1$A. Zawadowski, ${}^2$Jan von
    Delft and ${}^3$D. C. Ralph} \address{${}^{1}$Institute of Physics
    and Research Group of the Hungarian Academy of Sciences, Technical
    University of Budapest, \\ H-1521 Budafoki \'ut 8., Budapest,
    Hungary \\
    and Institute of Solid State Physics and Optics, H-1525 P.O. Box
    49,     Budapest, Hungary \\
    ${}^2$Institut f\"ur Theoretische Festk\"orperphysik,
    Universit\"at   Karlsruhe, 76128 Karlsruhe, Germany \\
    ${}^3$Laboratory of Atomic and Solid States Physics, Cornell
    University, Ithaca, New York 14853}

\date{Submitted on February 11, 1999; published in
 Phys. Rev. Lett {\bf 83}, 2632 (1999)}
\maketitle

\begin{abstract}
  We point out a novel, non-universal contribution to the dephasing
  rate $1/\tau_\varphi \equiv \gamma_\varphi$ of conduction electrons
  in metallic systems: scattering off non-magnetic two-level systems
  (TLSs) having almost degenerate Kondo ground states. In the regime
  $\Delta_{\rm ren} < T < T_K$ ($\Delta_{\rm ren}=$ renormalized level
  splitting, $T_K =$ Kondo temperature), such TLSs exhibit
  non-Fermi-liquid physics that can cause $\gamma_\varphi$, which
  generally decreases with decreasing $T$, to seemingly saturate in a
  limited temperature range before vanishing for $T\to 0$.  This could
  explain the saturation of dephasing recently observed in gold wires
  \mbox{[}Mohanty {\em et al.}\/ Phys. Rev. Lett. {\bf 78}, 3366
  (1997)\mbox{]}.
\end{abstract}
 \narrowtext

The dephasing behavior of conduction electrons in disordered systems
in the zero-temperature limit has recently been subject to
considerable and controversial discussions.  The standard theory of
dephasing in the context of weak localization \cite{Altshuler-Aronov},
predicts that the dephasing rate $1/\tau_\varphi \equiv
\gamma_\varphi$ (extracted from the magnetoresistance) vanishes for $T
\to 0$, since the phase space for inelastic scattering (e.g.\ 
electron-phonon or electron-electron) vanishes as the electron energy
approaches the Fermi energy.  In recent experiments on pure gold
wires, however, Mohanty, Jariwala and Webb (MJW) \cite{MJW97} found
that $\gamma_\varphi$ saturated at a {\em finite}\/ value for $T
\lesssim 1$K, though two common ``extrinsic'' sources of dephasing,
namely magnetic impurities and heating,
 were demonstrably absent.
Pointing out a similar saturation in older data on various other 1D
and 2D diffusive systems, MJW suggested \cite{MJW97} 
that the saturation could be due to a universal mechanism intrinsic to the
sample, namely ``zero-point fluctuations of phase coherent electrons''.
Although this suggestion contradicts the standard view,
Golubev and Zaikin \cite{GZ98lett} developed it into a detailed theory 
that was claimed to agree with 
numerous experiments.  However, their theory was criticized,
most strongly in \cite{AGA98}, but also in \cite{Danes,Vavilov}.
In \cite{AGA98} it was suggested that MJW's
elaborate shielding precautions were insufficient and
that  external microwave fields caused the saturation\vspace{-1mm}.

In this Letter, we reexamine another source of dephasing, non-universal but
intrinsic to any metal sample with structural disorder (which is never
completely absent), namely {\em dynamical two-level systems}\/ (TLSs), such as
point defects associated with dislocations, interfaces, surfaces or amorphous
regions.  TLSs were not considered as source of dephasing in the
above-mentioned debate (except very recently in \cite{Imry}), since 
standard inelastic scattering off non-degenerate
TLSs (assuming the standard uniform
distribution of level splittings, as discussed below)
gives $\gamma_\varphi \sim T$ \cite{BGJ79,Black}, which 
vanishes at low $T$.

Here, however,  we point out that 
another dephasing mechanism exists for TLSs in metals:
these are known to act as non-magnetic or orbital 2-channel Kondo
(2CK) impurities that exhibit non-Fermi-liquid (NFL) behavior in the
regime $\Delta_{\rm ren} < T < T_K$ ($\Delta_{\rm ren}=$
renormalized level splitting, $T_K =$ Kondo temperature), and we argue
below that {\em this NFL behavior includes dephasing};\/ in fact, it
can cause an apparent saturation in the decrease of $\gamma_\varphi$
with decreasing $T$ (although $\gamma_\varphi$ does tend to zero for
$T \to 0$).  This novel dephasing mechanism is {\em
  non-universal}\/, since the distribution of the material parameter
$T_K$ sets the energy scale, and since the density of TLSs depends on
the history of the sample. Reasonable assumptions for the density of
TLSs in Au lead to estimates for $\gamma_\varphi$ in accord with the
saturation behavior seen by MJW.

We start by noting that any dynamical impurity, i.e.\ one with
internal degrees of freedom, can potentially dephase
a conduction electron scattering off it: if in the process the
impurity changes its state, the electron's ``environment changes'',
and this, quite generally, causes dephasing \cite{SAI90}.
(In contrast, static impurities 
cannot change their state and hence cannot cause dephasing).

In this Letter, we focus on  dynamical ``spin 1/2'' impurities with two states,
denoted by $\Uparrow$ and $\Downarrow$,
which scatter free conduction electrons according to the rather general
interaction (specific examples are discussed below): 
\begin{equation}
  \label{eq:H-Kondo}
  H_I =  
\sum_{\varepsilon \varepsilon'} \sum_{\alpha \alpha' j \mu}
  c_{\varepsilon \alpha j}^\dagger 
   v^\mu_{\alpha \alpha'}
  c_{\varepsilon' \alpha' j} S_\mu \; . 
\end{equation}
The electrons are labeled by an energy $\varepsilon$, 
a ``spin'' index $\alpha$ that is not necessarily conserved
and a ``channel'' index $j$ that  (by definition) is conserved;
the $S_\mu$ ($\mu = x,y,z$) are 
spin-1/2 operators, with $S_z$ eigenvalues $(\Uparrow,\Downarrow) = ({1
  \over 2}, - {1\over 2})$; the coupling $ v^z$ describes
the difference in scattering potentials seen
by electrons scattering from the $\Uparrow$ or $\Downarrow$
state without flipping it, and is often called
a ``screening'' interaction, since it generates 
a ($S_z$-dependent) screening cloud around the impurity;
and $v^x$, $v^y$ describe scattering 
processes that ``flip the spin'' of the impurity.

{\em Magnetic Impurities:---}\/ 
As an illustrative example, let us briefly review dephasing
for the 1-channel Kondo model, for which
the channel index $j=1$ may be dropped and
$v^\mu_{\alpha \alpha'} = v \, \sigma^\mu_{\alpha \alpha'}$. 
Let $\gamma (T)$ be the
scattering rate of an electron at the Fermi surface
$(\varepsilon = 0)$;
it  can be split up as 
$ \gamma  = \gamma_\varphi + \gamma_{\rm pot}$ into parts that do or
do not cause dephasing, respectively (``pot'' for ``potential'' scattering).
Two kinds of processes contribute to $\gamma_\varphi$:
(i)  spin-flip scattering, as explained above, 
and (ii) single-to-many particle scattering
(see inset to Fig.~\ref{ss-sm}), 
since  additional electron-hole pairs can carry off phase information.
Figure~\ref{1ck}(a) shows the generic 
temperature-dependence of $\gamma$, $\gamma_{\rm pot}$
and $\gamma_\varphi$ \cite{Nozieres74}:
As $T$  approaches $T_K$ from above, 
all three rates increase logarithmically. As
$T$ is decreased past $T_K$, $\gamma$ continues to increase
monotonically, but crosses over to a
$(1- \mbox{const}\times T^2)$ behavior;
in contrast, $ \gamma_\varphi$ decreases
(this has been observed directly in the magnetoresistance of samples containing
magnetic impurities \cite{MJW97,Peters87}), since below $T_K$ the
formation of a Kondo singlet between the impurity and its screening
cloud begins to suppress spin-flip scattering.  For $T \ll T_K$
the singlet is inert (with spin-flip rate $\sim e^{-T/T_K}$),
and other conduction electrons experience only potential scattering
off it; they hence form a Fermi-liquid, in which a weak residual
interaction between electrons of opposite spins [Eq.~(D5) of
\cite{AL93}] yields a dephasing rate $\gamma_\varphi \propto
T^2/T^2_K$, which vanishes as $T \to 0$.


{\em TLSs in metals:---}\/ Next we consider 
an atom or group of atoms moving in  a double-well potential.
Labeling the separate ground states of the ``left'' and ``right'' well by 
 $(L,R) \equiv (\Uparrow, \Downarrow)$,
the bare Hamiltonian $\Delta_z  S_z + \Delta_x S_x$
describes a TLS with energy $\Delta_z$, 
spontaneous transition rate
$\Delta_x$ and level splitting $\Delta = \sqrt{\Delta_z^2 + \Delta_x^2}$
between the ground and excited states, say $ | \pm \rangle $. 
It is common \cite{Black} to assume a constant distribution
$P(\Delta) = \bar P$ of TLSs,
with  $\bar P\simeq 10^{19}- 10^{20} \, 
\mbox{eV}^{-1} \mbox{cm}^{-3}$ in metallic glasses.

When put in a metal, such a TLS will scatter conduction electrons.
The interaction's most general form is given by
Eq.~(\ref{eq:H-Kondo}), where now the ``spin'' index $\alpha$
classifies the electron's orbital state, representing e.g.\ its
angular momentum $(l,m)$, and the ``channel'' index $j = (\uparrow,
\downarrow)$ denotes its Pauli spin, which is conserved since the TLS
is non-magnetic.  This is in effect a generalized 2-channel Kondo
interaction, with which one can associate a Kondo temperature $T_K$.
In general it is highly anisotropic, with \linebreak
$|v^x|, |v^y| \ll |v^z|$,
since $v^x$, $ v^y$ describe electron-assisted 
inter-well transitions and depend on the barrier size
much more strongly than the screening interaction $v^z$ does.

{\em Slow fluctuators:---} If the barrier is sufficiently large ($|v^x|,
|v^y| \lll |v^z|$, so that $T_K \ll T, \Delta_x, \Delta_z$ and Kondo physics
is not important), the system is a ``slow fluctuator'', which can 
adequately be described by the so-called ``commutative'' model, in
which one sets $v^x = v^y = 0$ from the outset \cite{Black,Golding}.
This model does not renormalize to strong coupling at low
temperatures, and $\Delta$ is renormalized downward only slightly (by
at most a few \%) \cite{BVZ82}.  
%
%
To estimate $\gamma_\varphi$, one may thus use the bare parameters and
perturbation theory in $ v^z$, which couples  \cite{Black} $|+\rangle$ and $|-
\rangle$.
%
%
Since $v_z$-scattering between these, being inelastic, requires $T > \Delta$, 
the $\Delta$-averaged inelastic
scattering rate is $\overline \gamma_{\rm inel} \propto T$
(provided that $(\Delta_x)_{\rm max}, (\Delta_z)_{\rm max} >T$, 
\cite{BGJ79,Black}). 
 Thus $\overline \gamma_\varphi \propto T$ too, which 
does not saturate as $T\to 0$.

{\em Fast TLSs:---} For sufficiently small inter-well barriers,
however, the effective $T_K$ of a TLS can be significantly larger than
its effective level splitting, so that Kondo physics does come into
play \cite{estimateTK}. Such TLSs require the use of the full
``non-commutative'' model with $v^x$ and $v^y \neq 0 $, which flows
toward strong coupling under the renormalization group (RG)
\cite{VZ83,CZ98}.  Extensive RG studies  \cite{extensiveanalysis,Splittings}
showed that the regime $T\lesssim T_K$ is governed by an effective
isotropic 2CK interaction of the form (\ref{eq:H-Kondo}) with $\alpha
= (1,2)$ (since all but the two most-strongly-coupled orbital states
decouple) and $v^\mu_{\alpha \alpha'} = v \sigma^\mu_{\alpha
  \alpha'}$, and with an effective renormalized splitting $\Delta_{\rm ren} =
\Delta^2/T_K$.  In the so-called {\em NFL regime}\/ $\Delta_{\rm
  ren} < T < T_K$, the resulting effective 2CK model exhibits NFL
behavior \cite{AL93,AL91}. 
The zero-bias anomalies observed in recent years in nanoconstrictions
made from a number of different materials, such as Cu \cite{Ralph94},
Ti \cite{Upadhyay97} or metallic glasses \cite{Keijsers96,Balkashin},
can be consistently explained by attributing them to fast TLSs in or
near this 2CK NFL regime \cite{Annals1,WAMerratum}.  The Kondo
temperatures of the relevant TLSs were deduced in
\cite{Ralph94,Upadhyay97,Keijsers96} from the width of the zero-bias
anomalies to be $T_K \gtrsim 1$K, and in \cite{Balkashin} the
insensitivity of the anomalies to a high-frequency modulation of the
bias voltage implied $T_K \gtrsim 2$K.

{\em 2CK Dephasing:---}\/ Let us now consider dephasing due to fast
TLS in the NFL regime, a matter that to our knowledge has not been
addressed before.  In the NFL regime the single-to-single- and
single-to-many-particle scattering rates $\gamma_{\rm ss}(T)$
and $\gamma_{\rm sm} (T)$ (inset of Fig.~\ref{ss-sm}) are known to
respectively decrease and increase with decreasing $T$, in such a way
that the full scattering matrix is unitary \cite{unitarity}.  Now, our
key point is that {\em single-to-many-particle scattering must cause
  dephasing,}\/ so that we can take $\gamma_\varphi \simeq \gamma_{\rm
  sm}$. This immediately implies that the dephasing rate {\em
  increases with decreasing temperature}\/ in the NFL regime, as
indicated schematically in Fig.~\ref{2ck}(b). In fact, for $\Delta_{\rm ren} =
0$, one actually has $\gamma_{\rm ss} \propto (T/T_K)^{1/2} \to 0$ as
$T \to 0$ \cite{unitarity}, implying that the dephasing rate
($=\gamma_{\rm sm}$) would be finite even at $T=0$.  To heuristically
understand this result, recall that the NFL fixed point describes an
overscreened impurity that has a non-zero ground state entropy of ${1
  \over 2} \ln 2$ \cite{AL91} 
and cannot be viewed as an inert object (in contrast to 1CK case,
where the ground state singlet has entropy 0); intuitively speaking,
it is the dynamics associated with this residual entropy that causes
dephasing even at $T=0$.

Generally, however, $\Delta_{\rm ren} \neq 0$;
for $T < \Delta_{\rm ren} = \Delta^2/T_K$,  FL behavior is restored 
\cite{Splittings} and $\gamma_\varphi$ drops back to zero,
so that we crudely take $\gamma_\varphi (T) \simeq
 { \gamma_{\rm sm} (T) }
 \, \theta (\sqrt {T T_K} - \Delta)$.
Since NFL physics also requires $\Delta < T_K$,
we estimate the $\Delta$-averaged dephasing rate 
(with $P(\Delta) = \bar P $) as 
\begin{eqnarray}
\label{eq:averagedeph-T}
\overline   \gamma_\varphi^{T_K}  & \simeq &  \int_0^{T_K} d \Delta \, 
P(\Delta) { \gamma_\varphi (T) } \; ,
\end{eqnarray}
which yields $\overline \gamma_\varphi^{T_K} (T) \simeq \bar P \, {
  \gamma_{\rm sm} (T) } \, \mbox{min}[\sqrt {T T_K }, T_K]$. This has a broad
peak around $T_K$ [Fig.~1(c), dotted line].  
To next average over $T_K$, we
assume that the distribution $P( T_K)$ has
a broad maximum near, say, $\overline T_K$. 
Then the peak of $\overline \gamma_\varphi^{T_K} (T)$ would
be  broadened for $\overline \gamma_\varphi (T) = \int \! d T_K
P(T_K) \, \overline \gamma_\varphi^{T_K} (T)$ 
into a flattened region near $\overline T_K$.
Adding to this a power-law decay due to other sources of
dephasing, e.g.\ $\gamma_\varphi^{\rm 1D} \propto T^{2/3}$, the usual result
for disordered 1D wires \cite{Altshuler-Aronov}, the total dephasing rate
$\gamma_\varphi^{\rm tot} = \overline \gamma_\varphi + \gamma_\varphi^{\rm
  1D}$ would
have a broad shoulder 
 around $\overline T_K$, while vanishing for
$T\to 0$ [Fig.~1(c), solid line]. Thus {\em 2CK impurities can
  cause the total dephasing rate $\gamma_\varphi^{\rm tot}(T)$
to seemingly saturate in a limited
  temperature range.}\/ 

{\em Estimate of numbers:---} 
The shape of $\gamma_\varphi^{\rm tot} (T)$ and the 
existence of the broad shoulder depend on $P(T_K)$, $\overline
T_K$ and the relative weights of $ \overline \gamma_\varphi$ and
$\gamma_\varphi^{\rm 1D}$.  
To predict these from first principles would be overly ambitious,
since a microscopically reliable model for the TLSs and their couplings to
electrons is not available. Instead, let us use MJW's data to infer
what properties would be needed to attribute their saturation to 2CK
dephasing, and check the
inferred properties against other studies of TLSs.

The dephasing times in
Au wires saturated at $\tau_\varphi \simeq 5$ to 0.5ns below a crossover
temperature of about $T^\ast \simeq 1$K, which we 
associate with $\overline T_K$. We further assume 
the saturation to be dominated by 
TLSs with $\Delta_{\rm ren} < T^\ast < T_K$,
i.e.\ with $\Delta < {\rm 1K} < T_K$. 
Such parameters are reasonable, since experiment
\cite{Ralph94,Upadhyay97,Keijsers96,Balkashin} and theory
\cite{estimateTK} suggest that a sizable fraction of $\Delta<
1$K TLSs indeed do also have $T_K> 1$.
Let us estimate their required density. 
Impurities with dephasing
cross-section $\sigma_\varphi$ and density $n_i$ yield a dephasing rate
$\gamma_\varphi = v_F n_i \sigma_\varphi $.  
The density of strongly-coupled fast TLSs, i.e.\ 
with $\sigma_\varphi \lesssim \, \sigma_{\rm unit}$ close to the unitarity
limit $\sigma_{\rm unit} = 4 \pi/k_F^2$ per electron species, would thus have
to be of order $n_i = 1/(\tau_\varphi v_F \sigma_{\rm unit})$ $\gtrsim
2\times( 10^{15} -10^{16}) \mbox{cm}^{-3}$
(which is rather small: given the atomic density in Au of $6 \times
10^{22} \mbox{cm}^{-3}$, $n_i$ implies a 
TLS density of only $0.02 -0.2$~ppm \cite{highlystrained}).

The estimated value
 for $n_i$ is reasonable too: in metallic glasses,  the density of 
TLSs with splittings $\Delta < $1K is  $\bar P \times 1 \mbox{K}
\simeq 9 \times (10^{14} -10^{15})\mbox{cm}^{-3}$; in polycrystalline
Au, which is often taken to have roughly the same density of TLSs as
metallic glasses \cite{densityofTLS}, it is probably somewhat larger,
since (i) in polycrystals, which constitute a more symmetric
environment than glasses, the TLS distribution is probably more
heavily weighted for small splittings; (ii) in 1D wires, surface
defects can increase the total density of TLSs, and (iii) the bare
splittings $\Delta_z, \Delta_x$ are renormalized downward during the
flow toward the NFL regime \cite{Splittings}.  
The density of TLS in Au wires that can be expected
to cause 2CK dephasing thus compares satisfactorily with $n_i$ estimated
above. 


{\em Possible Checks:---} We emphasize that the 2CK dephasing
mechanism is non-universal: firstly, the energy scale is set
by $T_K$, and secondly, whether a sample contains sufficiently many
TLSs to cause appreciable dephasing depends on its history.
Thus, if the TLSs can be modified or even removed, e.g.\ by thermal
cycling or annealing \cite{Ralph94}, the dephasing behavior should
change significantly or even disappear.  Drawn wires containing more
dislocations (which may act as TLSs) should show stronger 2CK
dephasing than evaporated wires \cite{Sacharoff}.  Actually, already
in 1987 Lin and Giordano \cite{LG87} found hints in Au-Pd films of a
low-temperature dephasing mechanism that is ``very sensitive to
metallurgical properties''.  In semiconductors, however, TLSs are
unlikely to exhibit 2CK dephasing, since the much smaller
electron density implies much smaller couplings (for recent dephasing
experiments on semiconductors, see \cite{AGA98}).

In summary, we have pointed out a new, non-universal mechanism by
which two-level systems in metals, acting as 2CK impurities, can cause
dephasing, namely through an increased single-to-many-particle
scattering rate in their non-Fermi-liquid regime.  We estimate that
the Au wires of MJW \cite{MJW97} contain sufficiently many TLSs to
yield 2CK dephasing rates comparable to the saturation rates observed
there. More generally, though, the 2CK dephasing mechanism could be
used  to diagnose 2CK non-Fermi-liquid behavior in other
metals containing TLSs.

{\em Note added:---} Concurrent with this work, Imry,
Fukuyama and Schwab \cite{Imry}
proposed that $ 1/f$ noise from
 TLSs might produce essentially $T$-independent dephasing
by a different mechanism (not involving any 2CK physics), if 
$(\Delta_x)_{\rm max}$ is assumed to be $\ll T$ even for $T \simeq 1$K,
rather than the more common assumption \cite{BGJ79,Black}
that $\Delta_x$ has a larger range.

We thank I.\ Aleiner, B.\ Altshuler, V.\ Ambegaokar, Y.\ Imry, N.\
 Giordano, P.\ Nozi\`eres, H.\
Kroha, M.\  Vavilov, A.\ Zaikin, G.\  Zar\'and and W.\ Zwerger for discussions.
AZ benefited from the hospitality of the Meissner Institute and the LMU in
Munich; he was supported by the Hungarian Grants Nos.\ OTKA 96TO21228 and
97TO24005 and by the Humboldt Foundation, JvD by the DFG through SFB195, and
DCR by the MRSEC program of the NSF DMR-9632275 and the Packard Foundation.

\vspace{-1.1cm}.

\begin{figure}
\epsfxsize=0.9\linewidth
\epsfbox{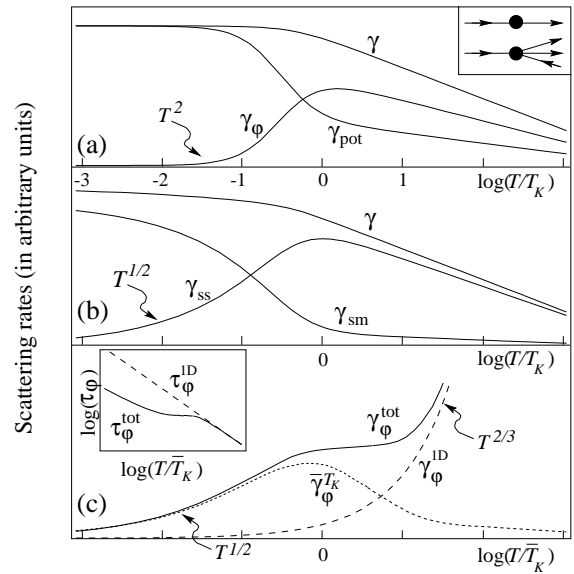}
\vspace{2mm}
\caption{
Scetch of scattering rates as functions of $\mbox{log}(T/T_K)$
for (a) the isotropic 1CK and (b) the anisotropic 2CK models
(for $\Delta_{\rm ren} = 0$);
inset: single-to-single- and
single\-to-many-particle scattering.
(c) Dotted line: a $\Delta$-averaged 2CK dephasing rate 
$\overline \gamma_\varphi^{\rm T_K}$ with $T_K \simeq \overline T_K$
(averaging such curves over $T_K$ yields $ \overline \gamma_\varphi$);
dashed line: 
$\gamma_\varphi^{\rm 1D} \sim T^{2/3}$;
full line: $\gamma_\varphi^{\rm tot} = 
\overline \gamma_\varphi + \gamma_\varphi^{\rm 1D}$;
inset: the corresponding dephasing times.}
\label{1ck} \label{ss-sm} \label{2ck}
\end{figure}
\noindent 

\widetext

\begin{thebibliography}{99}
\bibitem{estimateTK}
Recent theoretical estimates that
allow inter-well hopping via excited states gave $T_K$'s as large as
1-10K \cite{ZZ94,CZ98}.
\bibitem{Splittings} The RG flow of the initial anisotropic 2CK model
  has three stages: (i) as $T$ is lowered towards $T_K$, the couplings
  $v^\mu$ grow and become isotropic, $\Delta_z$ decreases only slightly
  and $\Delta_x$ decreases by two or more orders of magnitude, due to
  the orthogonality of the $L$ and $R$ electronic screening clouds
  \cite{VZ83} (for a recent discussion, see \cite{ZV97}, Fig.~6).  The
  effective splitting at $T \simeq T_K$ is thus roughly $\Delta
  \simeq (\Delta_z)_{\rm bare}$.  (ii) As $T$ decreases below $T_K$ the
  couplings increase further and the system flows towards a NFL fixed
  point. (iii) This flow is cut off once $T$ reaches the renormalized
  splitting $\Delta_{\rm ren} = \Delta^2/T_K$, below which the
  system flows towards a phase-shifted FL fixed point.
\bibitem{highlystrained}
Zero-bias anomalies in highly-strained nanoconstrictions suggest that
they contain much larger densities of strongly coupled 2CK impurities
of up to 100~ppm \cite{Ralph94}.
\end{thebibliography}
\end{document}